\documentclass[a4paper,12pt,oneside]{article}
\usepackage[latin1]{inputenc}
\usepackage[T1]{fontenc}
\usepackage[english]{babel}
\usepackage[mathscr]{euscript}
\usepackage{amsmath,exscale,latexsym,amssymb,amsthm,amsfonts}
\title{Notes on a particular Weyl Algebra}
\author{Giuseppe Iurato}
\date{}
\begin{document}\maketitle
\begin{abstract} By means of the notions of cross product algebras
of the theory of quantum groups, in the context of classical Hopf algebra structures, we deduce some known
structures of Weyl algebras type (as the Drinfeld quantum double, the restricted Heisenberg double, the
generalized Schr\"{o}dinger representation, and so on) that may be considered as a non-trivial examples of
quantum groups having physical meaning, starting from a particular example of groupoid motivated by elementary
quantum mechanics.\end{abstract}\large\bf 1. Introduction\\\\\rm\normalsize In the paper [Iu], following a
suggestion of Alain Connes (see [Co], I.1), it has been introduced a particular, simple groupoid, the so-called
\it Heisenberg-Born-Jordan EBB-groupoid \rm (or \it HBJ EBB-groupoid\rm), whose physical motivations were,
mainly, of spectroscopical nature.\\\\ An \it E-groupoid \rm (in the notations of [Iu]) is an algebraic system
of the type $(G, G^{(0)},r,s,\star)$, with $G,G^{(0)}$ non-void sets, $G^{(0)}\subseteq G$, $G^{(0)}$ set of
unities, $r,s:G\rightarrow G^{(0)}$ and $\star:G^{(2)}\rightarrow G$ partial groupoid law defined on
$G^{(2)}=\{(g_1,g_2)\in G\times G, s(g_1)=r(g_2)\}$, satisfying the set of axioms described in [Iu], § 1.\\\\
The HBJ EBB-groupoid is a particular E-groupoid that has been denoted with
$\mathcal{G}_{HBJ}(\mathcal{F}_I)=(\Delta\mathcal{F}_I,\mathcal{F}_I,r,s,\tilde{+})$, where
$\mathcal{F}_I=\{\nu_i;\nu_i\in\mathbb{R}^{+},i\in I\subseteq\mathbb{N}\}$ is the set of energy levels of a
certain spectroscopic physical system, $\Delta\mathcal{F}_I=\{\nu_{ij};\nu_{ij}=\nu_i-\nu_j,i,j\in
I\subseteq\mathbb{N}\}$, $r:(i,j)\rightarrow i$ and $s:(i,j)\rightarrow j$ are the range and source maps,
respectively, and $\nu_{ij}\tilde{+}\nu_{jk}=\nu_{ik}$ is the (partial) groupoid law as algebraic result of the
Ritz-Rydberg combination principle.\\\\ In [Iu], it has been only considered the structure of a no finitely
generated groupoid algebra on $\mathcal{G}_{HBJ}(\mathcal{F}_I)$, say
$\mathcal{A}_{\mathbb{K}}(\mathcal{G}_{HBJ}(\mathcal{F}_I))=(\langle\Delta(\mathcal{F}_I)\rangle,+,\cdot,\ast)$,
respect to an arbitrary commutative field $\mathbb{K}$ and a non-commutative convolution product $\ast$;
subsequently, it has been built up a (trivial) structure of braided non-commutative Hopf algebra on it.\\ We
claim that this last (albeit trivial) Hopf structure is the first and most natural possible one, on such a
groupoid algebra, because of the no (algebraic) finiteness of this generated algebra (since \it card $I=\infty$,
\rm in general).\\Therefore, the main interest of the paper [Iu], must be searched in the physical construction
of the EBJ EBB-groupoid.\\In this paper, we'll try to build up other (less trivial) structures on this special
HBJ EBB-groupoid, through adapted methods and tools of the theory of quantum groups, relative both to the
infinite-dimensional case and finite-dimensional case\footnote{The most interesting case, from the physical
view-point, is that finite-dimensional corresponding to \it card $I<\infty$, \rm since any physical
spectroscopic system has a finite number of energy levels.}.\\Furthermore, these structures will be introduced
taking into account eventual physical motivations.\\\\The (above mentioned) natural structure of Hopf algebra on
$\mathcal{A}_{\mathbb{K}}(\mathcal{G}_{HBJ}(\mathcal{F}_I))$ is given as follows: coproduct $\Delta(x)=x\otimes
x$, counit $\varepsilon(x)=1$, and antipode the extended inversion map.\\As already said, the first, natural
structure of a braided (or quasitriangular) non-commutative Hopf algebra on such an algebra, is trivially given
by the universal R-matrix $R=1\otimes 1$, whence a (trivial) example of quantum group if one assume a braided
(or quasitriangular) non-commutative Hopf algebra as definition of quantum group. Instead, a non-trivial example
of quasitriangular Hopf algebra arises from Drinfeld quantum double constructions (see § 6.).\\\\There exists
other definitions of a quantum group structure: for instance, if we consider a non-commutative and
non-cocommutative Hopf algebra as quantum group, then a cross (or bicross, or double cross) product construction
may provide examples of such a quantum group, whereas, if we consider as special 'quantum objects' the result of
a non-degenerate dual pairing of Hopf algebras, then a Heisenberg double may be taken as an example of quantum
group.\\\\If one want to determine examples of these last structures starting from
$\mathcal{G}_{HBJ}(\mathcal{F}_I)$, it is necessary, at first, examines the possible dual structures of
$\mathcal{A}_{\mathbb{K}}(\mathcal{G}_{HBJ}(\mathcal{F}_I))$, taking into account the existence of some problems
for this particular case study.\\ The first problem (that we'll sketch at the paragraph 3.) is related to
dualization in the infinite-dimensional case, whereas the second problem\footnote{In a certain sense,
preliminary to the first one.}, because of the infinity of $\mathcal{G}_{HBJ}(\mathcal{F}_I))$, is due to the
tentative of giving a Hopf algebra structure to the $\mathbb{K}$-algebra of $\mathbb{K}$-valued functions
defined on the HBJ EBB-groupoid $\mathcal{G}_{HBJ}(\mathcal{F}_I)$, say
$\mathcal{F}_{\mathbb{K}}(\mathcal{G}_{HBJ}(\mathcal{F}_I))$: in fact, on this algebra (that is strictly
correlated to the first problem of dualization of $\mathcal{A}_{\mathbb{K}}(\mathcal{G}_{HBJ}(\mathcal{F}_I))$,
and viceversa) it is a problematic question to define the right comultiplication and counit, for the following
reasons.\\\\For a group $(\mathcal{G},\cdot)$, the comultiplication question do not subsist in the
finite-dimensional case, because of the natural identification
$$\mathcal{F}_{\mathbb{K}}(\mathcal{G})\otimes\mathcal{F}_{\mathbb{K}}
(\mathcal{G})\cong\mathcal{F}_{\mathbb{K}}(\mathcal{G}\times\mathcal{G});$$in such a case, a natural structure
of Hopf algebra on $\mathcal{F}_{\mathbb{K}}(\mathcal{G})$, is given by the following data
:\begin{enumerate}\item coproduct
$\Delta:\mathcal{F}_{\mathbb{K}}(\mathcal{G})\rightarrow\mathcal{F}_{\mathbb{K}}(\mathcal{G}\times
\mathcal{G})$, given by $\Delta(f)(g_1,g_2)=f(g_1\cdot g_2)$, for all $g_1,g_2\in\mathcal{G}$;\item counit
$\varepsilon:\mathcal{F}_{\mathbb{K}}(\mathcal{G})\rightarrow\mathbb{K}$, with $\varepsilon(f)=1$;\item antipode
$S:\mathcal{F}_{\mathbb{K}}(\mathcal{G})\rightarrow\mathcal{F}_{\mathbb{K}}(\mathcal{G})$, defined as
$S(f)(g)=f(g^{-1})$ for all $g\in\mathcal{G}$,\end{enumerate}where the functional laws in 1. and 2. are
well-defined since, respectively, the group law is totally defined in $\mathcal{G}$, and there exists a unique
unit.\\\\Instead, if we consider a generic groupoid, these two questions remains unsolved, in the
finite-dimensional case too, both for the partial definition of the groupoid law and for the existence of many
unities: for these reasons, the initial definitions 1. and 2. of above, are ill-posed in this
case.\\\\Nevertheless, it is possible to solve these last problems with some extensions in the above
definitions, remaining in the context of classical Hopf algebra theory, but with minor usefulness of results.\\
Instead, in the new realm of the extended Hopf algebra structures, this problem may be clarified and solved with
fruitfulness, at least for the dual
$\mathcal{F}_{\mathbb{K}}^{\ast}(\mathcal{G}_{HBJ}(\mathcal{F}_I))\subseteq\mathcal{F}_{\mathbb{K}}
(\mathcal{G}_{HBJ}(\mathcal{F}_I))$.\\\\\large\bf 2. Cross product algebras\\\\\normalsize\rm The notions of
cross product and bicrossproduct are important tools for an algebraic setting of some fundamental structures of
Quantum Mechanics. In this paper, we'll to do only with the notion of cross (or smash, or semidirect)
product.\\\\Let $V_{\mathbb{K}}$ be a $\mathbb{K}$-linear space, $A$ a $\mathbb{K}$-algebra and $\psi:A\otimes
V\rightarrow V$ a $\mathbb{K}$-linear map; if we pose $\psi(h\otimes v)=\psi_h(v)$, and if
$\psi_{ab}(v)=\psi_a(\psi_b(v)),\ \ \psi_1(v)=v,\ \ \forall a,b\in A, \forall v\in V$, then
$(A,V_{\mathbb{K}},\psi)$ is a \it left $A$-module \rm on $V_{\mathbb{K}}$; we say that
$(A,V_{\mathbb{K}},\psi)$ is a \it left action \rm of $A$ on $V_{\mathbb{K}}$, or that $V_{\mathbb{K}}$ is a \it
left $A$-module \rm.\\Usually, we write $a\rhd v$ instead of $\psi_a(v)$, so that the action axioms are write as
$(ab)\rhd v=a\rhd(b\rhd v)$ and $1\rhd v=v$.\\If $A$ is a Hopf algebra, $V_{\mathbb{K}}$ is an $A$-module
algebra [coalgebra], and $a\rhd (vw)=(a_{(1)}\rhd v)(a_{(2)}\rhd w)$, $a\rhd 1_V=\varepsilon(h)1_V$
[$\Delta(a\rhd v)=(a_{(1)}\rhd v_{(1)})\otimes(a_{(2)}\rhd v_{(2)})$ (that is to say $\Delta(a\rhd
v)=\Delta_A(a)\rhd\Delta(v)$), $\varepsilon(a\rhd v)=\varepsilon(a)\varepsilon(v)$] for all $v,w\in
V_{\mathbb{K}}$ and $a\in A$, then $V_{\mathbb{K}}$ is said a \it left $A$-module algebra \rm[\it
coalgebra\rm].\\If $V$ is a Hopf algebra [bialgebra], then there exists the following two natural left actions
on itself: the \it left regular action $L$, \rm given by $L_v(w)=vw$, and the \it left adjoint action $Ad$, \rm
given by $Ad_v(w)=v_{(1)}wS(v_{(2)})$, for all $v,w\in V$.\\The \it left coregular action $R^{\ast}$ \rm of a
finite-dimensional Hopf algebra [bialgebra] $V$ on the dual $V^{\ast}$, is given by
$R^{\ast}_v(\phi)=\phi_{(1)}\langle v,\phi_{(2)}\rangle$, whereas, in the infinite-dimensional case, we set
$\langle R^{\ast}_v(\phi),w\rangle=\langle\phi,vw\rangle$, for all $v,w\in V$ and $\phi\in V^{\ast}$, being
$\langle\ , \ \rangle$ the dual pairing between $V$ and $V^{\ast}$ ($V^o$ in the infinite-dimensional case);
furthermore, $R^{\ast}$ makes $V^{\ast}$ [$V^o$] into a $V$-module algebra.\\The \it left coadjoint action \rm
of a finite-dimensional Hopf algebra [bialgebra] $V$ on the dual $V^{\ast}$, is given by
$Ad_v^{\ast}(\phi)=\phi_{(2)}\langle v,(S\phi_{(1)})\phi_{(2)}\rangle$, whereas, in the infinite-dimensional
case, we put $\langle Ad_v^{\ast}(\phi),w\rangle=\langle\phi,(Sv_{(1)})wv_{(2)}\rangle$, for all $v,w\in V$ and
$\phi\in V^{\ast}$, being $\langle\ ,\ \rangle$ the dual pairing between $V$ and $V^{\ast}$ ($V^o$ in the
infinite-dimensional case); furthermore, $Ad^{\ast}$ makes $V^{\ast}$ [$V^o$] into a $V$-module
coalgebra.\\\\The concept of $A$-module algebra generalizes\footnote{Because the structure of $A$-module (from
commutative algebra) generalize the notion of representation.} the notion of $G$-covariant algebra of the
Physics: if $G$ is a symmetry group, given a $G$-covariant $\mathbb{K}$-algebra $V$, we construct the group
algebra $\mathbb{K}G$ generated by $G$; then, the algebra generated by $\mathbb{K}G$ and $V$, with commutation
relations given by $uv=(u\rhd v)u\ \ \forall v\in V,u\in G$, give rise to a semidirect, or cross, product
algebra.\\ Therefore, we have the following general structure.\\Given a Hopf algebra [bialgebra] $A$ and a left
$A$-module algebra on $V$, then there exists a \it left cross product \rm algebra  on $V\otimes A$, with
product$$(v\otimes a)(w\otimes b)=v(a_{(1)}\rhd w)\otimes a_{(2)}b,\qquad v,w\in V,\ a,b\in A$$and unit element
$1\otimes 1$. This algebra is denoted with\footnote{Or with $V\rtimes_{\psi}A$, if one want to specify the
underling left action $\psi$.} $V\rtimes A$.\\\\With obvious modifications, it is possible to have right
actions, as follows.\\ A \it right action \rm of an algebra $A$ on the $\mathbb{K}$-linear space
$V_{\mathbb{K}}$, is a linear map $V\otimes A\rightarrow V$, denoted by $v\otimes a\rightarrow v\lhd a$, such
that $v\lhd(ab)=(v\lhd a)\lhd b$ and $v\lhd 1=v$ for all $a,b\in A,\ v\in V$. When $A$ is a Hopf algebra that
acts at right on an algebra $V$, and $(ab)\lhd v=(a\lhd v_{(1)})(b\lhd v_{(2)})$, $1_V\lhd v=1_V\varepsilon(v)$,
for all $a,b\in A$ and $v\in V$, then we say that $V$ is a \it right $A$-module algebra, \rm and we write
$(V,A,\lhd)$.\\A \it right $A$-coaction \rm on a $\mathbb{K}$-linear space $V_{\mathbb{K}}$, is a linear map
$\varphi:V\rightarrow V\otimes A$ such that $(\varphi\otimes\mbox{\rm id})\circ\varphi=(\mbox{\rm
id}\otimes\Delta)\circ\varphi$ and $\mbox{\rm id}=(\mbox{\rm id}\otimes\varepsilon)\circ\varphi$; we say, also,
that $V_{\mathbb{K}}$ is a \it right $A$-comodule.\rm \\If we set $\varphi(v)=v^{(\bar{1})}\otimes
v^{(\bar{2})}$ (in $V\otimes A$), then a coalgebra $V_{\mathbb{K}}$ is a \it right $A$-comodule \rm coalgebra if
$V_{\mathbb{K}}$ is a right $A$-comodule and$$v_{(1)}^{(\bar{1})}\otimes v_{(2)}^{(\bar{1})}\otimes
v^{(\bar{2})}=v_{(1)}^{(\bar{1})}\otimes v_{(2)}^{(\bar{1})}\otimes
v_{(1)}^{(\bar{2})}v_{(2)}^{(\bar{2})},\qquad\varepsilon(v^{(\bar{1})})v^{(\bar{2})}=\varepsilon(v).$$If $V$ is
a Hopf algebra [bialgebra], there are two natural right actions on itself: the \it right regular action R, \rm
given by $R_v(w)=wv$, and the \it right adjoint action Ad, \rm given by $Ad_v(w)=(Sv_{(1)})wv_{(2)}$, for all
$v,w\in V$.\\If $A$ is a Hopf algebra [bialgebra] and $V$ is a right $A$-comodule coalgebra, then there exists a
\it right cross coproduct \rm coalgebra structure on $A\otimes V$, given by$$\Delta(a\otimes v)=a_{(1)}\otimes
v_{(1)}^{(\bar{1})}\otimes a_{(2)}v_{(1)}^{(\bar{2})}\otimes v_{(2)},\quad\varepsilon(a\otimes
v)=\varepsilon_A(a)\varepsilon(c),$$for all $a\in A,\ v\in V$; such a coalgebra, is denoted with\footnote{Or
with $V\ltimes_{\psi}A$, if one want to specify the underling right action $\psi$.} $A\ltimes V$.\\\\
We do not discuss the notion of bicrossproduct algebra, introduced by S. Majid in connections with an important
tentative of unifying Quantum mechanics and Gravity (see [Ma], Chap. 6), but we give only a simple example of
what a bicrossproduct Hopf algebra is: let $G,M$ two subgroups that factorizes a given group, so that $G$ acts
on $M$, and viceversa (for instance, $M$ may be the position space, while $G$ may be the momentum group); let
$\mathbb{K}(M)$ the algebra of $\mathbb{K}$-valued functions on $M$, and let $\mathbb{K}G$ be the free algebra
on $G$. Then, the following Hopf algebra$$\mathbb{K}(M)\bowtie\mathbb{K}G=\left\{
\begin{array}{l}\mathbb{K}(M)\rtimes\mathbb{K}G\qquad\mbox{as\ algebra,}\\
\mathbb{K}(M)\ltimes\mathbb{K}G\qquad\mbox{as\ coalgebra}\end{array}\right.$$is a first example of
bicrossproduct algebra, whose dual Hopf algebra is $\mathbb{K}M\bowtie\mathbb{K}(G)$.\\Another notion strictly
correlated to that of bicrossproduct, is the notion of \it double cross product \rm (see [Ma]).\\\\The cross,
bicross and double cross product constructions, provides a large class of quantum groups.\\\\\large\bf 3. The
restricted Hopf algebra structure on
$\mathcal{F}_{\mathbb{K}}(\mathcal{G}_{HBJ}(\mathcal{F}_I))$\\\\\normalsize\rm There exists various methods to
define a classical Hopf algebra structure on $\mathcal{F}_{\mathbb{K}}(\mathcal{G}_{HBJ}(\mathcal{F}_I))$,
recalling that this algebra is infinite-dimensional.\\\\The main method, proceed as follows.\\In the case of the
groupoid $\mathcal{G}_{HBJ}(\mathcal{F}_I)$, that we recall to be a particular example of the general type
$(G,G^{(0)},r,s,\star)$, the most natural modifications to the functional laws on the points 1. and 2. of the §
1, are the following (see [Va], § 2.2):\begin{description}\item 1'. coproduct: $\Delta(f)(g_1,g_2)=f(g_1\star
g_2)$ if $(g_1,g_2)\in G^{(2)}$, and $=0$ otherwise;\item 2'. counit: $\varepsilon(f)=\sum_{e\in
G^{(0)}}f(e)$.\end{description}The antipode definition 3., is the same also in this case.\\\\Besides the
question relative to the functional laws, there exists the question related to their definition sets.\\Since, in
the infinite-dimensional case we have
$$\mathcal{F}_{\mathbb{K}}(\mathcal{G}_{HBJ}(\mathcal{F}_I))\otimes\mathcal{F}_{\mathbb{K}}(\mathcal{G}_{HBJ}
(\mathcal{F}_I))\subseteq\mathcal{F}_{\mathbb{K}}(\mathcal{G}_{HBJ}(\mathcal{F}_I)\times\mathcal{G}_{HBJ}
(\mathcal{F}_I))$$with$$\Delta:\mathcal{F}_{\mathbb{K}}(\mathcal{G}_{HBJ}(\mathcal{F}_I))\rightarrow
\mathcal{F}_{\mathbb{K}}(\mathcal{G}_{HBJ}(\mathcal{F}_I)\times\mathcal{G}_{HBJ}(\mathcal{F}_I)),$$let
$\mathcal{F}^{o}=\Delta^{-1}(\mathcal{F}_{\mathbb{K}}(\mathcal{G}_{HBJ}(\mathcal{F}_I))\otimes
\mathcal{F}_{\mathbb{K}}(\mathcal{G}_{HBJ}(\mathcal{F}_I)))
\subseteq\mathcal{F}_{\mathbb{K}}(\mathcal{G}_{HBJ}(\mathcal{F}_I))$; then, $\mathcal{F}^{o}$ is a Hopf algebra
with the coalgebra structure given by 1'., 2'. and 3., although it is difficult to determine exactly its
set-theoretic specificity.\\ It is called the \it restricted Hopf algebra \rm of
$\mathcal{F}_{\mathbb{K}}(\mathcal{G}_{HBJ}(\mathcal{F}_I))$, and is denoted with
$\mathcal{F}_{\mathbb{K}}^{o}(\mathcal{G}_{HBJ}(\mathcal{F}_I))$.\\\\For our purpose, in the finite-dimensional
case, there exists the isomorphism (see [Ks], III.1; [Ma], Example 1.5.4)
$\mathcal{F}_{\mathbb{K}}(\mathcal{G}_{HBJ}(\mathcal{F}_I))\cong\mathcal{A}_{\mathbb{K}}^{\ast}
(\mathcal{G}_{HBJ}(\mathcal{F}_I))$, so that we may construct a Hopf algebra structure on
$\mathcal{F}_{\mathbb{K}}(\mathcal{G}_{HBJ}(\mathcal{F}_I))$ via
$\mathcal{A}_{\mathbb{K}}^{\ast}(\mathcal{G}_{HBJ}(\mathcal{F}_I))$ by dual pairing; unfortunately, this
isomorphism do not subsists in the infinite-dimensional case, and such a question will be at the basis of the
discussion of § 6.\\\\Other methods for dualization (as Konstant duality, Cartier duality, Tannaka-Krein
duality, Takeuki duality, Kadison-Szlach\'{a}nyi dual pairing, the weak antipode plus convolution-inverse
method, Pontriyagin duality, and so on), may be found, for instance, in [Sw], [Sch1], [Sch2], [Sch3],
[Ma].\\\\However, in the context of the classical Hopf algebra structures, some of these methods do not lead to
an explicit solution of the problem, while others provides complicated structures unadapted to the physical
applications.\\But there exists different generalizations of the structure of Hopf algebra (for a recent survey
of these, see [Ka]) as, for instance, the notions of weak Hopf algebra (or quantum groupoid) and Hopf algebroid
(see [BNS], [NV]), through which it is possible to solve, more explicitly, the above problem, at least for the
dual $\mathcal{A}_{\mathbb{K}}^{\ast}(\mathcal{G}_{HBJ}(\mathcal{F}_I))$, in the context of weak Hopf algebras,
and with more possibilities on the side of physical applications.\\ Such a question, we'll be the matter of a
further paper.\\\\\large\bf 4. The restricted Heisenberg double
$\mathcal{H}^{o}_{\mathcal{A}}(\mathcal{G}_{HBJ}(\mathcal{F}_I))$\\\\\normalsize\rm Hence, as regard what has
been said above, we may consider the following dual pairing
$$\langle \ , \ \rangle:\mathcal{A}_{\mathbb{K}}(\mathcal{G}_{HBJ}(\mathcal{F}_I))\times\mathcal{F}_{\mathbb{K}}^{o}
(\mathcal{G}_{HBJ}(\mathcal{F}_I))\rightarrow\mathbb{K}$$such that$$\langle a_1\otimes
a_2,\Delta_{\mathcal{F}^{o}}(f)\rangle=\langle a_1a_2,f\rangle,\qquad\langle\Delta_{\mathcal{A}}(a), f_1\otimes
f_2\rangle=\langle a, f_1f_2\rangle$$$$\langle 1_{\mathcal{A}},f\rangle=\varepsilon_{\mathcal{F}^o}(f),\qquad \
\langle 1_{\mathcal{F}^o},a\rangle=\varepsilon_{\mathcal{A}}(a)$$for all $f,f_1,f_2\in\mathcal{F}^o,$ and
$a,a_1,a_2\in\mathcal{A}$.\\ It is know that it is always possible to consider, eventually quotienting, a
non-degenerate dual pairing of this type. Therefore, if we consider the action$$(b,a)\rightarrow b\rhd a=\langle
b,a_{(1)}\rangle a_{(2)}\qquad \forall a\in\mathcal{A}_{\mathbb{K}}(\mathcal{G}_{HBJ}(\mathcal{F}_I)),\ \forall
b\in\mathcal{F}_{\mathbb{K}}^o(\mathcal{G}_{HBJ}(\mathcal{F}_I)),$$it follows that it is possible to define the
left cross product algebra
$$\mathcal{H}_{\mathcal{A}_{\mathbb{K}}(\mathcal{G}_{HBJ}(\mathcal{F}_I)),
\mathcal{F}_{\mathbb{K}}^o(\mathcal{G}_{HBJ}(\mathcal{F}_I))}=
\mathcal{A}_{\mathbb{K}}(\mathcal{G}_{HBJ}(\mathcal{F}_I))\rtimes\mathcal{F}_{\mathbb{K}}^o
(\mathcal{G}_{HBJ}(\mathcal{F}_I)),$$called the \it Heisenberg double \rm of the pair
$\mathcal{A}_{\mathbb{K}}(\mathcal{G}_{HBJ}(\mathcal{F}_I),\mathcal{F}_{\mathbb{K}}^o(\mathcal{G}_{HBJ}
(\mathcal{F}_I))$. This last construction may be repeated for the restricted dual
$\mathcal{A}_{\mathbb{K}}^o(\mathcal{G}_{HBJ}(\mathcal{F}_I))(\subseteq\mathcal{A}_{\mathbb{K}}^{*}
(\mathcal{G}_{HBJ}(\mathcal{F}_I)))$ of $\mathcal{A}_{\mathbb{K}}(\mathcal{G}_{HBJ}(\mathcal{F}_I))$, obtaining
the so-called \it Heisenberg double \rm of $\mathcal{A}_{\mathbb{K}}(\mathcal{G}_{HBJ}(\mathcal{F}_I))$, that we
denote, for simplicity, with $\mathcal{H}_{\mathcal{A}}^o(\mathcal{G}_{HBJ}(\mathcal{F}_I))$.\\\\ \large\bf 5.
The restricted Weyl algebra\\\\\normalsize\rm The notion of cross product lead to an algebraic formulation of
some aspects of quantization.\\\\Let $V$ be a $A$-module algebra, with $A$ a Hopf algebra, and let $V\rtimes A$
be the corresponding left cross product. Hence, there exists a canonical representation on $V$ itself, given by
$(v\otimes a)\rhd w=v(a\rhd w)$, called the \it generalized Schr\"{o}dinger representation \rm of $V$.\\The
physical motivations to this terminology arise from the quantum meaning that such a representation has when
applied to the bicrossproduct algebra $\mathbb{K}(M)\rtimes\mathbb{K}G$ of the end of paragraph 2 (see also
[Ma], Chap. 6).\\\\Our interest is on infinite-dimensional case\footnote{But not only; for example, in the
finite-dimensional case, we have $V^o=V^{\ast}$, and what follows holds also in this case, with obvious
modifications.}, so let $V$ be a infinite-dimensional Hopf algebra, with restricted dual $V^o$; then, by the
left coregular representation $R^{\ast}$ of $V$ on $V^o$ (that does holds also in the infinite-dimensional case,
as seen at § 2.), $V^o$ is a $V$-module algebra, so that we may consider the left cross product algebra
$V^o\rtimes V$.\\\\Nevertheless, we are interested to another type of left cross product algebra, built up as
follows (see [Ma], § 6.1, for details).\\We consider the following action $\phi\rhd
v=v_{(1)}\langle\phi,v_{(2)}\rangle$ for all $v\in V,\phi\in V^o$, making $V$ into a $V^o$-module algebra and
that gives rise to the following product on $V\otimes V^o$
$$(v\otimes\phi)(w\otimes\psi)=vw_{(1)}\otimes\langle w_{(2)},\phi_{(1)}\rangle\phi_{(2)}\psi,$$whence a
structure of left cross product algebra on $V\otimes V^o$, namely $V\rtimes V^o$.\\Then, it is possible to prove
that the related Schr\"{o}dinger representation give rise to an isomorphism (of algebras) $\chi:V\rtimes
V^o\rightarrow Lin(V)$, where $Lin(V)$ is the algebra of $\mathbb{K}$-endomorphisms of $V$, given by
$\chi(v\otimes\psi)w=vw_{(1)}\langle\phi,w_{(2)}\rangle$.\\Therefore, we have the algebra isomorphism
$\mathcal{W}(V)=V\rtimes V^o\cong Lin(V)$; we call $\mathcal{W}(V)$ the \it restricted Weyl algebra \rm of the
Hopf algebra $V$.\\\\This last construction is an algebraic generalization of
the usual Weyl algebras of Quantum Mechanics on a group, whose finite-dimensional prototype is as follows.\\
We consider the strict dual pair given by the $\mathbb{K}$-valued functions on $G$, say $\mathbb{K}(G)$, and the
free algebra on $G$, say $\mathbb{K}G$; then, the left cross product algebra $\mathbb{K}(G)\rtimes\mathbb{K}G$
is given by the right action of $G$ on itself, namely $\psi_u(s)=su$, that induces a left regular representation
of $G$ on $\mathbb{K}G$, hence a Schr\"{o}dinger representation generated by this and by the action of
$\mathbb{K}G$ on itself by pointwise product. Whence, if $V=\mathbb{K}(G)$, we obtain the left cross product
algebra $\mathbb{K}(G)\rtimes\mathbb{K}G$, isomorphic to $Lin(\mathbb{K}(G))$ by Schr\"{o}dinger representation,
that formalizes the algebraic quantization of a particle moving on $G$ by translations.\\We may apply these
well-known considerations (see [Ma]) to $\mathcal{G}_{HBJ}(\mathcal{F}_I)$ when \it card $I<\infty$ \rm(finite
number of energy levels), taking into account that (in the finite-dimensional case)
$\mathcal{A}_{\mathbb{K}}(\mathcal{G}_{HBJ}(\mathcal{F}_I))=\mathcal{F}_{\mathbb{K}}^{\ast}
(\mathcal{G}_{HBJ}(\mathcal{F}_I))$, in such a way that the Weyl algebra
$$\mathcal{F}_{\mathbb{K}}(\mathcal{G}_{HBJ}(\mathcal{F}_I))\rtimes
\mathcal{A}_{\mathbb{K}}(\mathcal{G}_{HBJ}(\mathcal{F}_I))=\mathcal{F}_{\mathbb{K}}(\mathcal{G}_{HBJ}
(\mathcal{F}_I))\rtimes\mathcal{F}_{\mathbb{K}}^{\ast}(\mathcal{G}_{HBJ}(\mathcal{F}_I))=$$
$$=\mathcal{W}(\mathcal{F}_{\mathbb{K}}(\mathcal{G}_{HBJ}(\mathcal{F}_I)))\cong
Lin(\mathcal{F}_{\mathbb{K}}(\mathcal{G}_{HBJ}(\mathcal{F}_I)))$$represents the algebraic quantization of a
particle moving on the groupoid $\mathcal{G}_{HBJ}(\mathcal{F}_I)$ by translations\footnote{This remark may be
think as the starting point for a quantum mechanics on a groupoid.}.\\Instead, for the infinite-dimensional HBJ
EBB-groupoid $\mathcal{G}_{HBJ}(\mathcal{F}_I)$, we obtain a particular restricted Weyl algebra of the following
type
$$\mathcal{W}(\mathcal{G}_{HBJ}(\mathcal{F}_I))=\mathcal{F}_{\mathbb{K}}(\mathcal{G}_{HBJ}(\mathcal{F}_I))\rtimes
\mathcal{F}_{\mathbb{K}}^o(\mathcal{G}_{HBJ}(\mathcal{F}_I))$$with
$\mathcal{F}_{\mathbb{K}}^o(\mathcal{G}_{HBJ}(\mathcal{F}_I))\neq\mathcal{A}_{\mathbb{K}}
(\mathcal{G}_{HBJ}(\mathcal{F}_I))$ because of the no finitely generation of
$\mathcal{A}_{\mathbb{K}}(\mathcal{G}_{HBJ}(\mathcal{F}_I))$; therefore, the above physical interpretation of
the restricted Weyl algebra $\mathbb{K}(G)\rtimes\mathbb{K}G$, is no longer valid for
$\mathcal{W}(\mathcal{G}_{HBJ}(\mathcal{F}_I))$.\\However, as we'll see in another place, the last lack of
physical interpretation can be restored in the context of extended Hopf algebra structures.\\\\\large\bf 6. The
Drinfeld quantum double\\\\\normalsize\rm If $V$ is a Hopf algebra, then through the left adjoint action (in
infinite-dimensional setting) $Ad$ on itself, we have that $V$ is a $V$-module algebra, so that we may build the
left cross product algebra $V\rtimes_{Ad}V$.\\We consider the right adjoint action of
$\mathcal{G}_{HBJ}(\mathcal{F}_I)$ on itself, given by $\psi_g(h)=g^{-1}\star h\star g$ if exists, $=0$
otherwise; such an action makes $\mathcal{F}_{\mathbb{K}}(\mathcal{G}_{HBJ}(\mathcal{F}_I))$ into a
$\mathcal{A}_{\mathbb{K}}(\mathcal{G}_{HBJ}(\mathcal{F}_I))$-module algebra.\\In the finite-dimensional case we
have $\mathcal{A}_{\mathbb{K}}^{\ast}(\mathcal{G}_{HBJ}(\mathcal{F}_I))\cong
\mathcal{F}_{\mathbb{K}}(\mathcal{G}_{HBJ}(\mathcal{F}_I))$, hence
$$\mathcal{A}_{\mathbb{K}}(\mathcal{G}_{HBJ}(\mathcal{F}_I))=\mathcal{A}_{\mathbb{K}}^{\ast\ast}
(\mathcal{G}_{HBJ}(\mathcal{F}_I))\cong\mathcal{F}_{\mathbb{K}}^{\ast}(\mathcal{G}_{HBJ}(\mathcal{F}_I)),$$so
that $\mathcal{F}_{\mathbb{K}}(\mathcal{G}_{HBJ}(\mathcal{F}_I))$ is also a
$\mathcal{F}_{\mathbb{K}}^{\ast}(\mathcal{G}_{HBJ}(\mathcal{F}_I))$-module algebra, whence the left cross
product algebra$$\mathcal{F}_{\mathbb{K}}(\mathcal{G}_{HBJ}(\mathcal{F}_I))\rtimes
\mathcal{F}_{\mathbb{K}}^{\ast}(\mathcal{G}_{HBJ}(\mathcal{F}_I))\cong\mathcal{F}_{\mathbb{K}}
(\mathcal{G}_{HBJ}(\mathcal{F}_I))\rtimes\mathcal{A}_{\mathbb{K}}(\mathcal{G}_{HBJ}(\mathcal{F}_I))$$that the
tensor product coalgebra makes into a Hopf algebra, called the \it quantum double \rm of
$\mathcal{G}_{HBJ}(\mathcal{F}_I)$ and denoted with $D(\mathcal{G}_{HBJ}(\mathcal{F}_I))$; even in the
finite-dimensional case, it represent the algebraic quantization of a particle constrained to move on conjugacy
classes of $\mathcal{G}_{HBJ}(\mathcal{F}_I)$ (quantization on homogeneous spaces over a groupoid).\\Besides, it
was proved, for a finite group $G$, that this (Drinfeld) quantum double $D(G)$ has a quasitriangular structure
(see [Ma], Chap. 6), given by
$$(\delta_s\otimes u)(\delta_t\otimes v)=\delta_{u^{-1}su,t}\delta_t\otimes uv,\qquad \Delta(\delta_s\otimes u)=
\sum_{ab=s}\delta_a\otimes u\delta_b\otimes u,$$ $$\varepsilon(\delta_s\otimes u)=\delta_{s,e},\qquad
S(\delta_s\otimes u)=\delta_{u^{-1}s^{-1}u}\otimes u^{-1},$$ $$R=\sum_{u\in G}\delta_u\otimes e\otimes 1\otimes
u,$$where we have identifies the dual of $\mathbb{K}G$ with $\mathbb{K}(G)$ via the idempotents $p_g,\ g\in G$
such that $p_gp_h=\delta_{g,h}p_g$ (see [NV], 2.5. and [Ma], 1.5.4); such a quantum double represents the
algebra of quantum observables of a certain physical system.\\Hence, it is a natural question to ask if such
structures  may be extended to $D(\mathcal{G}_{HBJ}(\mathcal{F}_I))$, that is when we have a quantum double
built on a groupoid.\\\\\large\bf 7. Conclusions.\\\\\normalsize\rm From what has been said above (and in [Iu]),
it can be meaningful to study as the common known structures, just described above, may be extended when we
consider a groupoid\footnote{Finite or not.} instead of a finite group, both in the classical theory of Hopf
algebras and in the new realm of the extended Hopf structures.\\Subsequently, the resulting structures must be
interpreted from the physical view point, with a critical comparison respect to the physical meaning of the
classical Hopf structures just seen in this paper.\\These questions are not trivial, because there are recent
papers on a classical Physics on a groupoid (see, for instance, [CDMMM]), and many important works on the role
of Hopf algebras in High Energy Physics\footnote{Perhaps, it should be interesting to apply the extended Hopf
algebra structures to this context.} (see, for instance, [Kr] and references therein).\\\\\large\bf
References.\\\\\normalsize\rm {[BNS]} G. B\"{o}hm, F. Nill, K. Szlach\'{a}nyi, \it Weak Hopf Algebra I. Integral
Theory and $C^{\ast}$-structure, \rm J. Algebra, 221 (1999) pp. 385-438.\\{[Co]} A. Connes, \it Noncommutative
Geometry, \rm Academic Press, New York, 1994.\\{[CDMMM]} J. Cortes, M. De Leon, J.C. Marrero, D. Mart\'{i}n de
Diego, E. Mart\'{i}nez, \it A Survey of Lagrangian Mechanics and Control on Lie Algebroids and Groupoids, \rm
preprint e-arXiv: math.ph/0511009v1.\\{[Iu]} G. Iurato, \it A possible quantic motivation of the structure of
quantum group, \rm JP Journal of Algebra, Number Theory and Applications, 2010 (to appear).\\{[Ka]} G. Karaali,
\it On the Hopf algebras and their generalizations, \rm preprint e-arXiv: math.QA/0703441v2.\\{[Ks]} C. Kassel,
\it Quantum Groups, \rm Springer-Verlag, New York, 1995.\\{[Kr]} D. Kreimer, \it Algebraic structures in Quantum
Field Theory, \rm preprint e-arXiv: hep-th/1007.0341v1.\\ {[Ma]} S. Majid, \it Foundations of quantum group
theory, \rm Cambridge University Press, Cambridge, 1995.\\{[NV]} D. Nikshych, L. Vainerman, \it Finite quantum
groupoids and their applications, \rm in: \it New directions in Hopf algebras, \rm S. Montgomery, H.J. Schneider
eds., MSRI Publications, Vol. 43, (2002), pp. 211-262.\\{[Sch1]} P. Schauenburg, \it Tannaka Duality for
arbitrary Hopf Algebras, \rm Verlag Reinhard Fischer, Munich, 1992.\\{[Sch2]} P. Schauenburg, \it Duals and
doubles of quantum groupoids ($\times_R$-bialgebras), \rm in: \it New Trends in Hopf Algebra Theory, \rm N.
Andruskiewitsch, W.R. Ferrer Santos, and H.J. Schneider, Eds., vol. 267, Amer. Math. Soc., New York,
2000.\\{[Sch3]} P. Schauenburg, \it Weak Hopf Algebras and Quantum Groupoids, \rm preprint e-arXiv:
math.QA/0204180v1.\\{[Sw]} M. Sweedler, \it Hopf Algebras, \rm W.A. Benjamin Inc., New York, 1969.\\{[Va]} J.M.
Vallin, \it Relative matched pairs of finite groups from depth two inclusions of Von Neumann algebras to quantum
groupoid, \rm preprint e-arXiv: math.OA/0703886v1.

\end{document}